\documentclass[10pt,twocolumn]{ICCAS}

\usepackage{amsmath}
\usepackage{amssymb}
\usepackage{diagbox}
\usepackage{bm}
\usepackage{xcolor}

\DeclareSymbolFont{letters}{OML}{cmm}{m}{it}
\DeclareSymbolFont{operators}{OT1}{cmr}{m}{n}
\DeclareSymbolFont{symbols}{OMS}{cmsy}{m}{n}
\DeclareSymbolFont{largesymbols}{OMX}{cmex}{m}{n}

\newcommand{\Enc}{\operatorname{Enc}}
\newcommand{\Dec}{\operatorname{Dec}}
\newcommand{\Pack}{\operatorname{Pack}}
\newcommand{\UnpackPt}{\operatorname{UnpackPt}}

\newcommand{\col}{\operatorname{col}}
\newcommand{\Range}{\operatorname{Range}}
\newcommand{\rank}{\operatorname{rank}}

\newtheorem{thm1}{\bf Theorem}
\newtheorem{prop1}{\bf Proposition}
\newtheorem{lem1}{\bf Lemma}
\newtheorem{asm1}{\bf Assumption}
\newtheorem{defn1}{\bf Definition}
\newtheorem{rem1}{\bf Remark}
\newtheorem{cor1}{\bf Corollary}

\newenvironment{asm}{\begin{asm1}}{\hfill$\square$\end{asm1}}
\newenvironment{rem}{\begin{rem1}}{\hfill$\square$\end{rem1}}

\newcommand{\Gcal}{{\mathcal{G}}}
\newcommand{\Rcal}{{\mathcal{R}}}
\newcommand{\Zbb}{{\mathbb{Z}}}
\newcommand{\Nbb}{{\mathbb{N}}}
\newcommand{\Rbb}{{\mathbb{R}}}

\newcommand{\rscale}{{\mathrm{r}}}
\newcommand{\sscale}{{\mathrm{s}}}
\newcommand{\Lscale}{{\mathrm{L}}}

\newcommand{\taufb}{\tau_{\mathrm{fb}}}
\newcommand{\tauilc}{\tau_{\mathrm{ilc}}}

\begin{document}

% \title{Ring-LWE Based Encrypted Iterative Learning Control with Error Accumulation}
\title{On Controlling the Effect of Error Growth in Unlimited Encrypted Iterative Learning Control}

\author{Sangwon Lee${}^{1}$ and Junsoo Kim${}^{1*}$ }

\affils{ ${}^{1}$Department of Electrical and Information Engineering, Seoul National University of Science and Technology,\\
Seoul, 01811, Korea (leesangwon@cdslst.kr, junsookim@seoultech.ac.kr)}

\thanks{
This work was supported by the National Research Foundation of Korea (NRF)
grant funded by the Korea government (MSIT) (No. RS-2024-00353032).
}

\abstract{
This paper proposes a Ring Learning With Errors (Ring-LWE) based encrypted iterative learning control (ILC) framework for repetitive tracking tasks over networked control systems. The architecture integrates an encrypted dynamic feedback controller with an encrypted ILC computation. During each trial, the feedback controller is evaluated in the ciphertext domain, and the encrypted output trajectory is stored directly in the cloud. 
After each trial, the cloud evaluates the tracking error and performs the 
ILC computation from the stored ciphertexts, so that the plant
side does not need to store the accumulated trial data.
The proposed framework uses distinct packing parameters for ciphertext multiplication, allowing the cloud to handle both lower-dimensional output feedback control and higher-dimensional ILC computation without decryption. 
While error growth in Ring-LWE based encrypted control is generally suppressed by closed-loop stability, the marginally stable ILC iterations cause the injected errors to accumulate continuously.
%In addition,% 
To address this challenge, a range-space decomposition is introduced in the encrypted ILC formulation to allow evaluation under unlimited updates.
Numerical simulations show that the range-space decomposition suppresses
encryption-induced perturbation, while ciphertext packing improves the computational efficiency of the encrypted ILC update.
}


\maketitle

%-----------------------------------------------------------------------
\section{Introduction}
\label{sec:introduction}

Cloud platforms and remote computing have become common in networked control
systems, but they may expose control signals and system information to
untrusted cyber-environments. Disclosed signals and parameters may reveal
sensitive information and enable cyber-attacks \cite{Sandberg15CSM}.
Homomorphic encryption addresses this issue by allowing control computations
to be performed directly on encrypted signals without decrypting the
underlying data \cite{Kogiso15CDC,Kim22ARC}.
Encrypted feedback control schemes have been implemented in the ciphertext
domain \cite{Kim23TAC,Teranishi23TCNS}, and homomorphic encryption
has also been applied to data-driven control and controller tuning
\cite{Alexandru20CDC,Hoshino25SICE}. These results show that
data-dependent control computations can be outsourced while plant 
signals are protected.

Iterative learning control (ILC) improves tracking performance for systems
that repeatedly perform the same finite-duration task by updating feedforward
inputs from previous trial data \cite{Owens05ARC}. A secure ILC method has been
studied to protect trial signals against disclosure attacks \cite{Chen24TCAS}.
However, existing approaches mainly focus on signal privacy, while control or
learning parameters may still be visible to the computing side. Therefore,
these parameters also need to be protected when the learning computation is
outsourced to an external computing platform.

Ring Learning With Errors (Ring-LWE) \cite{Lyubashevsky13JACM,Chillotti16ASIACRYPT} based homomorphic encryption provides an
efficient framework for encrypted computation over polynomial rings. 
%However,
Still, ciphertext multiplication remains a costly operation. Packing methods alleviate this cost by processing multiple messages within a single ciphertext, and they have been used in Ring-LWE based encrypted control methods \cite{Lee25SMCS,Jang25TCNS}.
Since ILC requires handling large-sized encrypted data and evaluating ciphertext multiplications, packing is essential to make such computation practical.
If ``encrypted ILC'' is realizable, cloud-side storage of output ciphertexts
can reduce the plant-side burden of accumulating trial data. 

%homomorphic multiplications such as leveled
%Ring-LWE multiplication and the Ring-GSW (Gentry-Sahai-Waters) external

This paper proposes an encrypted ILC framework based on Ring-LWE that combines
the encrypted dynamic feedback controller in \cite{Jang25TCNS} with an encrypted ILC
computation. 
The encrypted computations are outsourced to a cloud that operates over ciphertexts without access to the secret key.
The cloud stores encrypted output trajectories and computes the feedforward update directly over encrypted data, while the plant side only receives the updated feedforward input for the next trial.
The proposed framework deals with two major issues:
\begin{itemize}
    \item In Ring-LWE based encrypted control, encryption-induced errors arise naturally, and the resulting error growth under recursive operations can be suppressed by stability of the closed-loop system \cite{Jang25TCNS}.
    In contrast, since ILC iterations are generally marginally stable, a suitable mechanism will be needed to prevent error accumulation.
    \item The stored output
ciphertexts should be usable in both the feedback and ILC computations, in which
different packing parameters should be used.
\end{itemize}

To resolve these issues, we propose to use
component-wise encrypted output trajectories to share the stored ciphertexts
for both the feedback and the ILC computations, and introduce a range-space
decomposition to suppress null-space error accumulation under unlimited
encrypted feedforward updates.
%Although the injected errors in the marginally stable component of the encrypted ILC iteration accumulate in an uncontrolled manner, their effect can be eliminated through a projection operation in the computation of the ILC feedforward input.
In particular, the ILC iteration dynamics is reduced to eliminate the marginally stable component while preserving the original feedforward update, thereby resolving the error accumulation problem.
Simulation results will demonstrate the computational benefits of the proposed method and its effectiveness in controlling the effect of error growth.
%\colb{We validate the proposed encrypted ILC framework by numerical simulation, confirming the effect of the range-space decomposition and the computational benefit of ciphertext packing.}

%In addition, repeated feedforward updates can accumulate encryption-induced error in the encrypted feedforward input. These considerations will be important when combining encrypted feedback control with encrypted ILC.

The remainder of this paper is organized as follows.
Section 2 introduces the encryption scheme and problem
setting. Section 3 presents the proposed encrypted ILC
framework. Section 4 provides simulation results,
and Section 5 concludes the paper.

\textit{Notation:}
The sets of integers, positive integers, and real
numbers are denoted by $\Zbb$, $\Nbb$, and $\Rbb$, respectively. 
The floor, ceiling, and rounding operations are denoted by
\(\lfloor\cdot\rfloor\), \(\lceil\cdot\rceil\), and
\(\lceil\cdot\rfloor\), respectively.  
For \(q\in\Nbb\), let 
\(\Zbb_q:=\Zbb\cap[-q/2,q/2)\), and let \(a\bmod q\) denote the modulo operation defined by $
    a \bmod q
    :=
    a-\left\lfloor {(a+q/2)}/{q}\right\rfloor q
    \in \Zbb_q .
$
For vectors \(a_0,\ldots,a_{n-{1}}\),
\(\col\{a_0,\ldots,a_{n-{1}}\}\) denotes column-wise stacking. 
The identity matrix of size \(n\) is denoted by \(I_n\).
The Kronecker product is denoted by \(\otimes\). 
For a vector or matrix, \(\|\cdot\|\) denotes the infinity
norm or the induced infinity norm.
For a polynomial
\(a(X)=\sum_{i=0}^{N-{1}}a_iX^i\), define
\(\|a\|:=\max_{0\leq i<N}|a_i|\). 
The ILC iteration index is denoted by \(j\), while the discrete-time index within one trial is denoted by \(k\).

%-----------------------------------------------------------------------
\section{PRELIMINARIES AND PROBLEM SETTING}
\label{sec:preliminaries}

\subsection{Ring-LWE Encryption Scheme}
% This subsection summarizes the Ring-LWE and Ring-GSW notation and packing
% operations used in this paper.
Let \(N\in\Nbb\) be a power of two and \(q\in\Nbb\) be a prime modulus. 
Let \(\Rcal_q:=\Zbb_q[X]/\langle X^N+1\rangle\) be the ring of integer polynomials
modulo \(X^N+1\),
%and \(\Rcal_q\) denote its residue ring modulo \(q\),
whose coefficients belong to %elements are polynomials in \(\Rcal\) with coefficients in
\(\Zbb_q\).
Throughout this paper, ciphertexts are written in bold type.
For a plaintext \(m\in \Rcal_q\), the Ring-LWE encryption is defined as
\begin{equation*}
    \Enc(m)
    :=
    \begin{bmatrix}
        sk\cdot a+m+e\\
        a
    \end{bmatrix}
    \mod q \in \Rcal^2_q,
\end{equation*}
where \(sk\in \Rcal_q\) is the secret key whose coefficients are sampled
uniformly at random from \(\{-1,0,1\}\), \(a\) is sampled uniformly at random from
\(\Rcal_q\), and \(e\in\Rcal_q\) is sampled from a truncated discrete Gaussian
distribution \(\psi\) over \(\Rcal_q\) satisfying \(\|e\|\leq\sigma_{\rm RLWE}\).
For a ciphertext ${\mathbf{c}} = \Enc(m)$, the decryption is defined as
\begin{equation*}
    \Dec({\mathbf{c}})
    :=
    \begin{bmatrix}
        {1} & -sk
    \end{bmatrix}
    {\mathbf{c}}
    =
    m+e
    \mod q.
\end{equation*}

For ciphertexts \({\mathbf{c}}_1\in\Rcal_q^2\) and \({\mathbf{c}}_2\in\Rcal_q^2\),
the homomorphic addition and subtraction are defined as
\begin{align*}
    {\mathbf{c}}_1\boxplus{\mathbf{c}}_2 &:= {\mathbf{c}}_1+{\mathbf{c}}_2 \mod q \in \Rcal^2_q, \\
    {\mathbf{c}}_1\boxminus{\mathbf{c}}_2 &:= {\mathbf{c}}_1-{\mathbf{c}}_2 \mod q \in \Rcal^2_q,
\end{align*}
and they satisfy $\Dec({\mathbf{c}}_1\boxplus{\mathbf{c}}_2)=\Dec({\mathbf{c}}_1)+\Dec({\mathbf{c}}_2)$
and $\Dec({\mathbf{c}}_1\boxminus{\mathbf{c}}_2)=\Dec({\mathbf{c}}_1)-\Dec({\mathbf{c}}_2)$.
In practice, one uses a message-recovery parameter \(\Lscale\) with
$1/\Lscale\in\Nbb$. 
For a message $m$, the scaled encryption and decryption are
\begin{align*}
    \Enc_{\Lscale}(m)
    &:=
    \Enc(m/{\Lscale} \mod q),\\
    \Dec_{\Lscale}({\mathbf{c}})
    &:=
    \left\lceil
        \Lscale \cdot \Dec({\mathbf{c}})
    \right\rfloor
    \mod q.
\end{align*}
If $1/\Lscale > 2\sigma_{\mathrm{RLWE}}$, 
then $\Dec_{\Lscale}(\Enc_{\Lscale}(m)) = m$ holds for all $m \in \Rcal_q$ 
with $\|m\| < \Lscale q/2 - 1/2$.

Ring-GSW (Gentry--Sahai--Waters) encryption is used to perform ciphertext multiplication through the
external product \cite{Chillotti16ASIACRYPT}. Let \(\nu\in\Nbb\) be the %decomposition
base, \(d=\lceil\log_\nu q\rceil\), and define
\(G_{\nu} = \begin{bmatrix}{1} & \nu & \cdots & \nu^{d-{1}}\end{bmatrix}\otimes I_2\in \Zbb^{2\times 2d}\).
For \(M\in \Rcal_q\), a Ring-GSW ciphertext is written as
\begin{equation*}
    \Enc'(M)
    :=
    PM\cdot G_{\nu}+Z
    \mod qP \in \Rcal_{qP}^{2\times 2d},
\end{equation*}
where \(P\in\Nbb\) is the ``special modulus'' for the external product, and the
columns of \(Z\in \Rcal_{q}^{2\times 2d}\) are Ring-LWE encryptions of zero.
The external product
\(\boxdot:\Rcal_{qP}^{2\times 2d}\times\Rcal_{q}^{2}\to\Rcal_{q}^{2}\)
between a Ring-GSW and a Ring-LWE ciphertext satisfies
\begin{equation*}
\Dec(\Enc'(M)\boxdot{\mathbf{c}})=M\Dec({\mathbf{c}})+\Delta \mod q    
\end{equation*}
for some bounded \(\Delta\in\Rcal_q\), with details in \cite{Jang25TCNS}.
For vectors and matrices of ciphertexts, \(\boxplus\) and \(\boxminus\) are
applied component-wise, and \(\boxdot\) follows the matrix-vector convention,
with the resulting external products summed via \(\boxplus\).

The packing and unpacking operations are summarized as follows.
For a power of two $\tau$ with $n\leq\tau\leq N$ and
$\alpha=[\alpha_0,\ldots,\alpha_{n-1}]^\top\in\Zbb_q^n$,
\begin{equation*}
    \Pack_{n}(\alpha):=\sum_{i=0}^{n-1}\alpha_i X^{iN/\tau}\in \Rcal_q,
\end{equation*}
and for $p(X)=\sum_{i=0}^{N-1}p_iX^i\in \Rcal_q$,
\begin{equation*}
    \UnpackPt_{n}(p)
    :=
    \col\{p_0,p_{N/\tau},\ldots,p_{(n-1)N/\tau}\}.
\end{equation*}

The detailed Ring-LWE/Ring-GSW cryptosystem, packing and unpacking algorithms,
and error analysis are given in \cite{Jang25TCNS}.

\subsection{Problem Setting}
Consider a discrete-time linear plant repeatedly performing the same
finite-duration tracking task of length \(T\in\Nbb\) over
\(k=0,1,\ldots,T-1\) as
\begin{equation}\label{eq:linear_plant}
\begin{aligned}
    x_j(k+{1}) &= Ax_j(k)+Bu_j(k), \quad x_j(0)=x_0,\\
    y_j(k) &= Cx_j(k),
\end{aligned}    
\end{equation}

where \(j\) is the iteration index, \(x_j(k)\in\Rbb^{n_x}\) is the state with the initial value \(x_0\in\Rbb^{n_x}\),
\(u_j(k)\in\Rbb^m\) is the input, and \(y_j(k)\in\Rbb^p\) is the output.
%\(A\in\Rbb^{n_x\times n_x}\), \(B\in\Rbb^{n_x\times m}\),
%\(C\in\Rbb^{p\times n_x}\), and \(x_0\in\Rbb^{n_x}\) is the initial state.
For each $j$-th trial, an observer-based dynamic feedback controller driven by
the reference \(r(k)\in\Rbb^p\) is designed as 
%\begin{equation}
%\begin{aligned}
%    z_j(k+{1}) &= Fz_j(k)+Gy_j(k)+Wr(k), \, z_j(0)=z_0,\\
%    u^{\rm fb}_j(k) &= Hz_j(k)+Jr(k),
%\end{aligned}
%\label{eq:feedback}
%\end{equation}
\begin{align}
    z_j(k+{1}) &= Fz_j(k)+Gy_j(k)+Wr(k), \, z_j(0)=z_0,\notag\\
    u^{\rm fb}_j(k) &= Hz_j(k)+Jr(k),\label{eq:feedback}
\end{align}
where \(z_j(k)\in\Rbb^{n_z}\) is the state with the initial value \(z_0\in\Rbb^{n_z}\), and \(u^{\rm fb}_j(k)\in\Rbb^m\) is the controller output.
%\(F\in\Rbb^{n_z\times n_z}\), \(G\in\Rbb^{n_z\times p}\),
%\(H\in\Rbb^{m\times n_z}\), \(J\in\Rbb^{m\times p}\),
%\(W\in\Rbb^{n_z\times p}\), and \(z_0\in\Rbb^{n_z}\) is the initial controller state.
The condition \(F\in\Zbb^{n_z\times n_z}\) is required for encrypted
recursive implementation \cite{Cheon18CDC}, and the transformed
controller satisfying this condition is assumed to be given \cite{Kim23TAC}.

In addition to the feedback input, an ILC feedforward
input \(v_j(k)\in\Rbb^m\) is applied as
\begin{equation*}
    u_j(k)=u^{\rm fb}_j(k)+v_j(k).
\end{equation*}
We describe how to iteratively update and design $v_j(k)$, for $j=1,2,\ldots$. Define
%the closed-loop state
\(X_j(k):=\col\{x_j(k),z_j(k)\}\in\Rbb^{n_x+n_z}\).
%with the initial value \(X_0:=\col\{x_0,z_0\}\).
The initial state \(X_0:=\col\{x_0,z_0\}\) and the reference \(\{r(k)\}_{k=0}^{T-1}\) are fixed across all trials.
The closed-loop dynamics are written as
\begin{equation}
\begin{aligned}
    X_j(k+1) &= A_{\rm cl}X_j(k)+B_{\rm cl}v_j(k)+B_r r(k),\\
    y_j(k) &= C_{\rm cl}X_j(k),
\end{aligned}
\label{eq:cl_dynamics}
\end{equation}
where
\begin{align*}
    A_{\rm cl}
    &=
    \begin{bmatrix}
        A & BH \\
        GC & F
    \end{bmatrix},
    \quad
    B_{\rm cl}
    =
    \begin{bmatrix}
        B \\ 0
    \end{bmatrix},
    \quad
    B_r
    =
    \begin{bmatrix}
        BJ \\ W
    \end{bmatrix},
    \\
    C_{\rm cl}
    &=
    \begin{bmatrix}
        C & 0
    \end{bmatrix}.
\end{align*}
The lifted signals are defined as
\begin{align*}
Y_j &:= \col\{y_j(0),\ldots,y_j(T-1)\} \in \Rbb^{pT},\\
V_j &:= \col\{v_j(0),\ldots,v_j(T-1)\} \in \Rbb^{mT},\\
R   &:= \col\{r(0),\ldots,r(T-1)\} \in \Rbb^{pT}.
\end{align*}
Then, the lifted input-output relation is represented as
\begin{equation}
    Y_j = \Phi X_0 + \Gcal_r R + \Gcal V_j,
    \label{eq:lifted}
\end{equation}
where \(\Phi\in\Rbb^{pT\times(n_x+n_z)}\) and
\(\Gcal\in\Rbb^{pT\times mT}\) are given by
{\small
\begin{equation*}
    \Phi
    :=
    \begin{bmatrix}
        C_{\rm cl} \\
        C_{\rm cl}A_{\rm cl} \\
        \vdots \\
        C_{\rm cl}A_{\rm cl}^{T-1}
    \end{bmatrix},
    \, \Gcal
    :=
    \begin{bmatrix}
        0 & \cdots & \cdots & 0 \\
        C_{\rm cl}B_{\rm cl} & \ddots & & \vdots \\
        \vdots & \ddots & \ddots & \vdots \\
        C_{\rm cl}A_{\rm cl}^{T-2}B_{\rm cl} & \cdots & C_{\rm cl}B_{\rm cl} & 0
    \end{bmatrix},
\end{equation*}}
\!\!and \(\Gcal_r\in\Rbb^{pT\times pT}\) is defined analogously
to \(\Gcal\) with \(B_{\rm cl}\) replaced by \(B_r\).
Let $Y^{\rm fb} := \Phi X_0 + \Gcal_r R$
denote the baseline output trajectory, which is fixed across all trials.
Then, \eqref{eq:lifted} simplifies to
\begin{equation}
    Y_j = Y^{\rm fb} + \Gcal V_j.
    \label{eq:ILC_law}
\end{equation}

\begin{asm} \label{asm:rangeG}
For all iterations, \eqref{eq:ILC_law} holds, and the
compensation target \(R^* := R - Y^{\rm fb}\) satisfies
\begin{equation*}
    R^* \in \Range(\Gcal).
\end{equation*}    
\end{asm}

The objective is then to drive the tracking error
\(E_j := R - Y_j \in \Rbb^{pT}\) to zero over iterations while keeping all
trial data encrypted in the ciphertext domain.
%cloud.%

%-----------------------------------------------------------------------
\section{Main Result}
\subsection{Feedback Controller with Trajectory Storage}

The implementation of the feedback part \eqref{eq:feedback} directly follows \cite[Section~IV]{Jang25TCNS}, where the
columns of the controller matrices are packed and encrypted as Ring-GSW
ciphertexts for efficient encrypted dynamic control.
Let \(F_i\), \(G_i\),
\(W_i\), \(H_i\), \(J_i\) denote the \(i\)-th columns of $F$, $G$, $W$, $H$, $J$ respectively,
and 
choose a power of two
\(\taufb\) with \(\max\{n_z,m,p\}\leq\taufb\leq N\).
The packed Ring-GSW ciphertexts are encrypted as
\begin{align*}
    {\mathbf{F}}_i &= \Enc'\left(\Pack_{n_z}(F_i \bmod q)\right),
    \quad i=1,\ldots,n_z,\\
    {\mathbf{G}}_i &= \Enc'\left(\Pack_{n_z}\left(\left\lceil G_i/\sscale\right\rfloor \bmod q\right)\right),
    \quad i=1,\ldots,p,\\
    {\mathbf{W}}_i &= \Enc'\left(\Pack_{n_z}\left(\left\lceil W_i/\sscale\right\rfloor \bmod q\right)\right),
    \quad i=1,\ldots,p,\\
    {\mathbf{H}}_i &= \Enc'\left(\Pack_{m}\left(\left\lceil H_i/\sscale\right\rfloor \bmod q\right)\right),
    \quad i=1,\ldots,n_z,\\
    {\mathbf{J}}_i &= \Enc'\left(\Pack_{m}\left(\left\lceil J_i/\sscale\right\rfloor \bmod q\right)\right),
    \quad i=1,\ldots,p,
\end{align*}
where $1/\sscale\ge 1$ is the scale factor for matrices.
The signals $y_j(k)$ and $r(k)$ are encrypted component-wise, as
\begin{align*}
    {\mathbf{y}}_{j,i}(k)
    &=
    \Enc\left(
        \left\lceil y_{j,i}(k)/\rscale \right\rfloor
        /\Lscale
        \mod q
    \right), \\
    {\mathbf{r}}_{i}(k)
    &=
    \Enc\left(
        \left\lceil r_i(k)/\rscale \right\rfloor
        /\Lscale
        \mod q
    \right),
\end{align*}
where \(y_{j,i}(k)\) and \(r_i(k)\) denote the \(i\)-th components of
\(y_j(k)\) and \(r(k)\), respectively, for \(i=1,\ldots,p\), $\rscale>0$ is the quantization step size, and $1/\Lscale\in\Nbb$ is the
message-recovery scale factor.
The initial controller state is encrypted as
\begin{equation*}
    {\mathbf{z}_0}
    :=
    \Enc\left(
        \Pack_{n_z}
        \left(
            \left\lceil z_0/(\rscale\sscale) \right\rfloor
        \right)
        /\Lscale
        \mod q
    \right).
\end{equation*}

The cloud constructs the encrypted trajectories
\begin{align*}
{\mathbf{Y}}_j
&:=
\col\{
{\mathbf{y}}_{j,1}(0),\ldots,{\mathbf{y}}_{j,p}(0),\\
&\hspace{1.2cm}
\ldots,
{\mathbf{y}}_{j,1}(T-{1}),\ldots,{\mathbf{y}}_{j,p}(T-{1})
\},\\
{\mathbf{R}}
&:=
\col\{
{\mathbf{r}}_{1}(0),\ldots,{\mathbf{r}}_{p}(0),\\
&\hspace{1.2cm}
\ldots,
{\mathbf{r}}_{1}(T-{1}),\ldots,{\mathbf{r}}_{p}(T-{1})
\},
\end{align*}
from the component-wise ciphertexts, storing \({\mathbf{Y}}_j\) during each
trial and preparing \({\mathbf{R}}\) in advance.
The encrypted tracking error ${\mathbf{E}}_j:={\mathbf{R}} \boxminus {\mathbf{Y}}_j$ 
is then computed and used for the encrypted ILC update without decrypting the tracking signals.

\begin{rem}\label{rem:err_analysis}
The error analysis in \cite{Jang25TCNS} assumes a fixed packing parameter,
where the output is packed before encryption and unpacked in the ciphertext
domain for the controller computation. In the proposed framework, however,
the stored output is shared between the feedback and ILC computations, which
use different packing parameters. Unpacking a packed output ciphertext and
reusing it under a different parameter would mix the encryption errors across
packing slots. The output and reference trajectories are therefore encrypted
component-wise as Ring-LWE ciphertexts rather than packed.
\end{rem}

\subsection{ILC Update and Error Accumulation}
\begin{figure}[thb]
\begin{center}
\includegraphics[width=8.0cm]{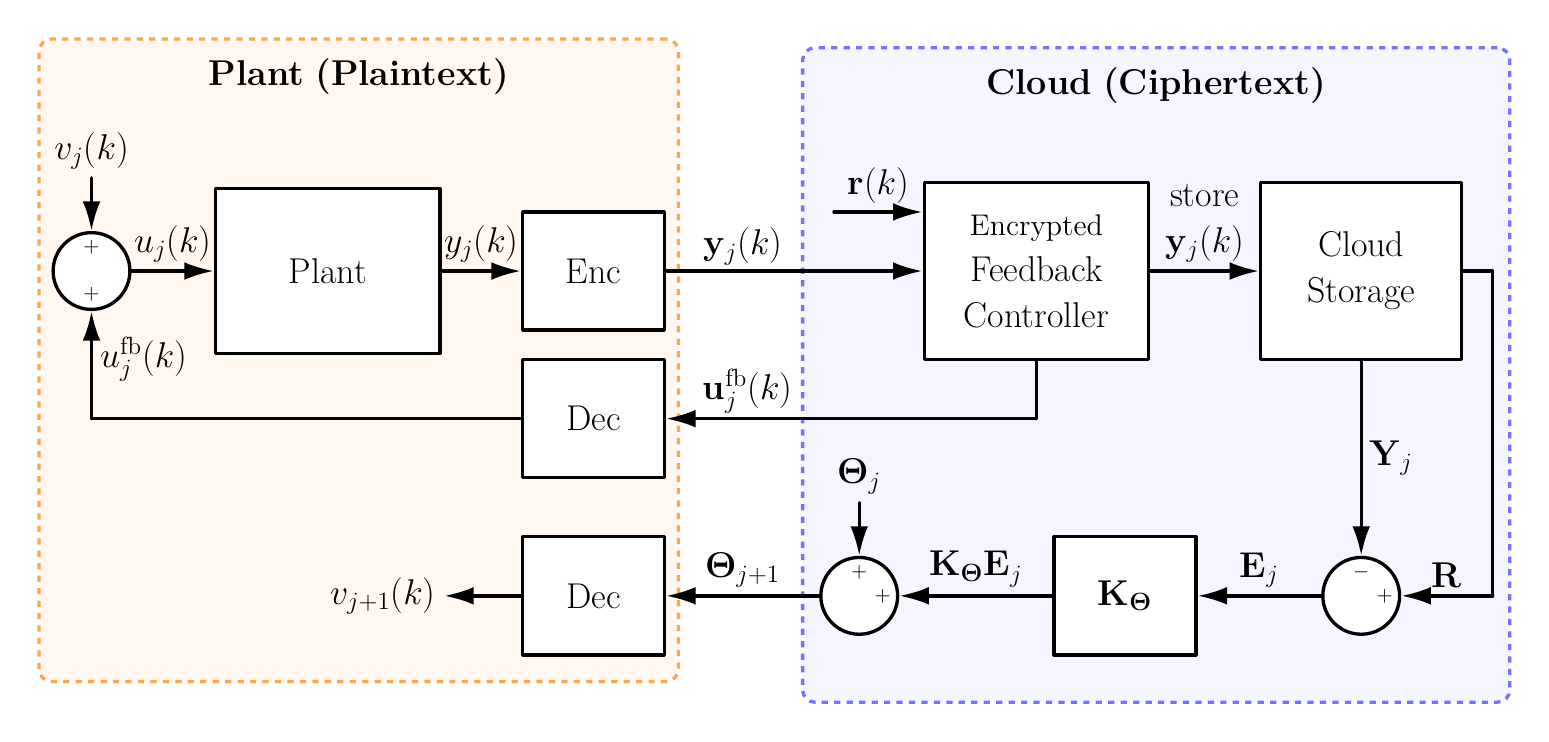}
\caption{\label{fig:encrypted_ilc_architecture} Diagram of the proposed encrypted ILC architecture with encrypted dynamic feedback controller.}
\end{center}
\end{figure}

We first describe the nominal ILC update. 
Let $K \in \Rbb^{mT \times pT}$ be the learning gain matrix.
The ILC update rule is given by
\begin{equation}
    V_{j+1}=V_j+K E_j,
    \label{eq:full_ilc}
\end{equation}
where the tracking error is derived as
\begin{equation}
    E_j = R-Y_j = R^*-\Gcal V_j.
    \label{eq:tracking_error}
\end{equation}
Following the adjoint algorithm of \cite{Owens05ARC}, we choose the learning gain $K$ with the step size $\eta$, as
\begin{equation}
    K=\eta\Gcal^\top,
    \qquad
    0<\eta<
    \frac{2}{\lambda_{\max}(\Gcal^\top\Gcal)},
    \label{eq:full_gain}
\end{equation}
where \(\lambda_{\max}(\cdot)\) denotes the largest eigenvalue.
From \eqref{eq:full_ilc} and \eqref{eq:tracking_error}, the tracking error dynamics can be represented as
\begin{multline*}
    E_{j+1}
    = R^* - \Gcal V_{j+1} 
    = R^* - \Gcal (V_j + KE_j) \\
    = E_j - \Gcal KE_j 
    = (I-\eta\Gcal\Gcal^\top)E_j .
\end{multline*}
Under Assumption~\ref{asm:rangeG}, $R^* \in \Range(\Gcal)$, and since
$\Gcal V_j \in \Range(\Gcal)$, the error $E_j = R^* - \Gcal V_j$ also belongs to $\Range(\Gcal)$ for all $j$.
Then, the design \eqref{eq:full_gain} ensures that
the mapping $I-\eta\Gcal\Gcal^\top$ restricted to the domain $\Range(\Gcal)$ is Schur
stable \cite{Owens05ARC}, so $E_j$ converges to zero as $j$ increases.

When \eqref{eq:full_ilc} is performed in encrypted form using
the stored ciphertext trajectories, let $\widetilde{V}_j$ denote
the feedforward input obtained after decryption.
Analogously to the error-growth analysis of \cite{Jang25TCNS}, the decrypted
update can be modeled as
\begin{equation}
    \widetilde V_{j+1}
    =
    \widetilde V_j+ K E_j+d_j,
    \label{eq:perturbed_ilc}
\end{equation}
where $d_j \in \Rbb^{mT}$ is the encryption-induced perturbation. 
Substituting $E_j = R^* - \Gcal\widetilde{V}_j$ into \eqref{eq:perturbed_ilc} with \eqref{eq:full_gain} gives
\begin{equation*}
    \widetilde V_{j+1}
    =
    (I-\eta\Gcal^\top\Gcal)\widetilde V_j
    +
    \eta\Gcal^\top R^*
    +
    d_j .
\end{equation*}
Since $I-\eta\Gcal^\top\Gcal$ has unit eigenvalues in $\ker(\Gcal)$,
the $\ker(\Gcal)$-components of $d_j$ accumulate unboundedly in
$\widetilde{V}_j$, eventually violating the plaintext range constraint.

To avoid this null-space accumulation, the feedforward input is restricted to
\(\Range(\Gcal^\top)\). Let
\begin{equation*}
    n_\Theta:=\rank(\Gcal),\qquad
    S\in\Rbb^{mT\times n_\Theta},
\end{equation*}
where the columns of \(S\) form an orthonormal basis of
\(\Range(\Gcal^\top)\), satisfying
\begin{equation*}
    \Range(S)=\Range(\Gcal^\top),
    \qquad
    S^\top S=I_{n_\Theta}.
\end{equation*}
The feedforward input is represented by the reduced coordinate
\(\Theta_j\in\Rbb^{n_\Theta}\) as
\begin{equation*}
    V_j=S\Theta_j,\qquad
    \Gcal_\Theta:=\Gcal S,\qquad
    \Theta_0=S^\top V_0 .
\end{equation*}
Then the lifted relation becomes
\begin{equation*}
    Y_j=Y^{\rm fb}+\Gcal_\Theta\Theta_j .
\end{equation*}
Since \(S\) spans \(\Range(\Gcal^\top)\),
\(\Gcal_\Theta\in\Rbb^{pT\times n_\Theta}\) has full column rank and
\(\Range(\Gcal_\Theta)=\Range(\Gcal)\).
Choose the reduced learning gain $K_\Theta$ with the step size $\gamma$ as
\begin{equation}
    K_\Theta
    =
    \gamma\Gcal_\Theta^\top,
    \qquad
    0<\gamma<
    \frac{2}{\lambda_{\max}(\Gcal_\Theta^\top\Gcal_\Theta)} .
    \label{eq:reduced_gain}
\end{equation}
Let $\widetilde{\Theta}_j$ denote the reduced coordinate obtained after
decryption.
% , with $Y_j = Y^{\rm fb} + \Gcal_\Theta\widetilde{\Theta}_j$.
The reduced-coordinate ILC update after decryption is derived as
\begin{equation}
    \widetilde{\Theta}_{j+1}
    =
    \widetilde{\Theta}_j+K_\Theta E_j + \delta_j,
    \label{eq:reduced_ilc}
\end{equation}
where $\delta_j\in\Rbb^{n_\Theta}$ is the encryption-induced perturbation.
Substituting $E_j = R^* - \Gcal_\Theta\widetilde{\Theta}_j$ into \eqref{eq:reduced_ilc} with \eqref{eq:reduced_gain} gives
\begin{equation*}
    \widetilde{\Theta}_{j+1}
    =
    (I-\gamma\Gcal_\Theta^\top\Gcal_\Theta)\widetilde{\Theta}_j
    +\gamma\Gcal_\Theta^\top R^*
    +\delta_j.
\end{equation*}
By \eqref{eq:reduced_gain} and the full column rank of $\Gcal_\Theta$,
$I-\gamma\Gcal_\Theta^\top\Gcal_\Theta$ is Schur stable.
Following \cite{Jang25TCNS}, where closed-loop stability suppresses
encryption-induced perturbation in recursive encrypted control,
the stability of the reduced-coordinate ILC update suppresses
the bounded perturbation $\delta_j$ in
$\widetilde{\Theta}_j$ for all $j$.
The resulting behavior is evaluated numerically in Section~4.

\subsection{Packing Implementation} 

We now turn to the encrypted implementation of the reduced-coordinate ILC update, where the dominant cost lies in the external products. 
To quantify the benefit of packing, we first present the non-packed implementation and then its packed counterpart.
Let
\(\boldsymbol{\kappa}_\Theta\) denote the non-packed Ring-GSW encryption of
\(K_\Theta\), where each element is encrypted separately as
\begin{equation*}
    \boldsymbol{\kappa}_\Theta
    =
    \Enc'\left(
        \left\lceil
            K_\Theta/\sscale
        \right\rfloor
    \mod q
    \right)
    \in
    \left(\Rcal_{qP}^{2\times 2d}\right)^{n_\Theta\times pT}.
\end{equation*}
Let \(\boldsymbol{\xi}_j\) denote the non-packed Ring-LWE ciphertext vector
corresponding to \(\Theta_j\), where each component is encrypted
separately, initialized by
\begin{equation*}
    \boldsymbol{\xi}_0
    =
    \Enc\left(
        \left\lceil
            \Theta_0/(\rscale\sscale)
        \right\rfloor
        /\Lscale
        \mod q
    \right)\in
    \left(\Rcal_q^2\right)^{n_\Theta}.
\end{equation*}
The corresponding encrypted ILC update is
\begin{equation}
    \boldsymbol{\xi}_{j+1}
    =
    \boldsymbol{\xi}_{j}
    \boxplus
    \left(
        \boldsymbol{\kappa}_\Theta
        \boxdot
        {\mathbf{E}}_j
    \right),
    \label{eq:nonpack}
\end{equation}
which requires \(n_\Theta pT\) external products.

Next, we apply the column-wise packing method of \cite{Jang25TCNS}. Choose a power
of two \(\tauilc\) satisfying \(n_\Theta\leq\tauilc\leq N\). For
\(\ell=1,\ldots,pT\), let \(K_{\Theta,\ell}\) be the \(\ell\)-th column of
\(K_\Theta\), and let \({\mathbf{E}}_{j,\ell}\in\Rcal_q^2\) denote the
\(\ell\)-th component-wise ciphertext of \({\mathbf{E}}_j\). Define
\begin{equation*}
    {\mathbf{K}}_{\Theta,\ell}
    =
    \Enc'\left(
        \Pack_{n_\Theta}
        \left(
            \left\lceil
                K_{\Theta,\ell}/\sscale
            \right\rfloor
            \mod q
        \right)
    \right).
\end{equation*}
The initial reduced coordinate $\boldsymbol{\Theta}_0$ is initialized as a packed Ring-LWE ciphertext as
\begin{equation*}
    {\boldsymbol{\Theta}}_0
    =
    \Enc\left(
        \Pack_{n_\Theta}
        \left(
            \left\lceil
                \Theta_0/(\rscale\sscale)
            \right\rfloor
        \right)
        /\Lscale
        \mod q
    \right).
\end{equation*}
The packed encrypted ILC update is
\begin{equation}
    {\boldsymbol{\Theta}}_{j+1}
    =
    {\boldsymbol{\Theta}}_j
    \boxplus
    \left(
    \sum_{\ell=1}^{pT}
    {\mathbf{K}}_{\Theta,\ell}
    \boxdot
    {\mathbf{E}}_{j,\ell}
    \right),
    \label{eq:packed}
\end{equation}
where the summation is taken with respect to \(\boxplus\). After the encrypted
update, the plant side decrypts, unpacks, rescales, and maps the reduced
coordinate back to the feedforward input, as
\begin{equation*}
    V_{j+1}
    =
    \rscale\sscale\Lscale\,
    S\cdot
    \UnpackPt_{n_\Theta}
    \left(
        \Dec({\boldsymbol{\Theta}}_{j+1})
    \right).
\end{equation*}
The non-packed update \eqref{eq:nonpack} requires
\(n_\Theta pT\) external products, whereas the packed update
\eqref{eq:packed} requires only \(pT\). Thus, column-wise packing reduces the
number of external products by a factor of \(n_\Theta\).

%-----------------------------------------------------------------------
\section{Simulation Results}
\label{sec:simulation_results}

%This section verifies the proposed encrypted ILC framework using
For a numerical
%discrete-time
example, let the system be given as \eqref{eq:linear_plant}
% \begin{align*}
%     x_j(k+1)&=Ax_j(k)+Bu_j(k), \qquad x_j(0)=x_0,\\
%     y_j(k)&=Cx_j(k),
% \end{align*}
where \(x_j(k)\in\Rbb^4\), \(u_j(k)\in\Rbb^2\), \(y_j(k)\in\Rbb\), and $x_0 = 0$.
The system matrices of the plant model are
\begin{equation*}
A=
\begin{bmatrix}
 0.9972 & 0.0497 & 0      & 0      \\
-0.1117 & 0.9853 & 0      & 0      \\
 0      & 0      & 0.9804 & 0.0485 \\
 0      & 0      &-0.7759 & 0.9338
\end{bmatrix},
\end{equation*}
\begin{equation*}
B=
\begin{bmatrix}
0.0124 & 0.0050\\
0.4965 & 0.1986\\
0.0037 & 0.0123\\
0.1455 & 0.4849
\end{bmatrix},
\quad
C=
\begin{bmatrix}
1 & 0 & 0.5 & 0
\end{bmatrix}.
\end{equation*}
Each trial lasts \(10{\rm s}\), with sampling period \(T_s=0.05{\rm s}\), and
hence \(T=200\).

The observer-based controller \eqref{eq:feedback} is designed by
discrete-time LQR with
\(Q_{\rm lqr}=\operatorname{diag}(10,1,10,1)\) and \(R_{\rm lqr}=I_2\),
yielding the feedback gain \(K_{\rm lqr}\in\Rbb^{m\times n_x}\),
and the observer is placed at poles \(\{0.70,0.73,0.76,0.79\}\),
with \(z_j(0)=0\) for all trials.
The reference gain matrices are chosen as
\(J = N_{\rm bar}\) and \(W = BN_{\rm bar}\),
where \(N_{\rm bar}\in\Rbb^{m\times p}\) is the DC gain satisfying
\(C(I-A+BK_{\rm lqr})^{-1}BN_{\rm bar}=I_p\).
The controller is transformed to have integer state matrix \(F\in\Zbb^{n_z\times n_z}\)
following \cite{Kim23TAC}.

The reference trajectory, fixed over all trials, is
\[
    r(k)
    =
    \sin^2\!\left(\frac{\pi k}{200}\right)
    \left\{
        \sin\!\left(\frac{3\pi k}{200}\right)
        -
        0.5\sin\!\left(\frac{2\pi k}{200}\right)
    \right\}.
\]
The lifted dimensions are \(mT=400\) and \(pT=200\). The reduced basis
\(S\) is computed as an orthonormal basis of \(\Range(\Gcal^\top)\), giving
\(n_\Theta=199\). The learning rates are set to $\eta=0.01$ and $\gamma=0.01$,
satisfying \eqref{eq:full_gain} and \eqref{eq:reduced_gain} with
$\lambda_{\max}(\Gcal^\top\Gcal) = \lambda_{\max}(\Gcal_\Theta^\top \Gcal_\Theta) = 99.42$,
and the initial feedforward input is $V_0 = 0$.

To ensure 128-bit security \cite{Albrecht21HE}, the Ring-LWE/Ring-GSW
parameters are set to \(N=2^{12}\), \(\log q=56\), and \(\log P=51\),
with the error distribution \(\psi\) set as a discrete Gaussian of standard deviation \(3.2\)
bounded by \(\sigma_{\rm RLWE}=19.2\). The scaling factors are
\(\rscale=\sscale=10^{-3}\) and \(\Lscale=10^{-2}\), and the feedback
packing parameter is \(\taufb=4\) and the ILC packing parameter is
\(\tauilc=512\). Simulations were performed using Lattigo
v6.1.0 \cite{Lattigo24} on a desktop PC with an AMD Ryzen 5 7500F CPU at
3.70 GHz and 32 GB RAM.

\begin{figure}[!t]
\centering
\includegraphics[width=8.0cm]{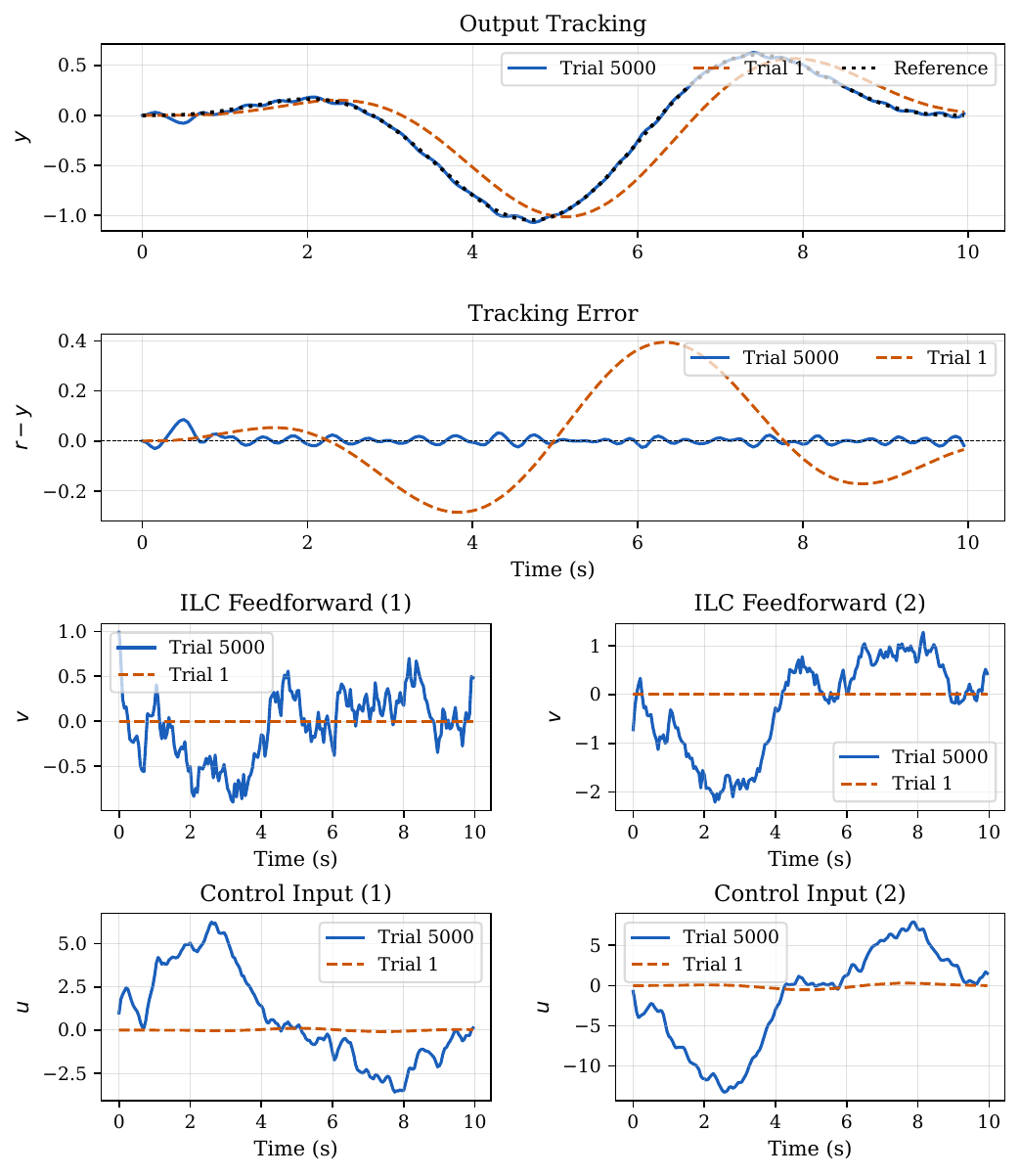}
\caption{\label{fig:without_reduction}
Encrypted ILC result for the full-coordinate update without range-space
reduction. The reference trajectory (dotted black line),
initial trial (dashed orange line), and final trial (solid blue line) are shown.}

\includegraphics[width=8.0cm]{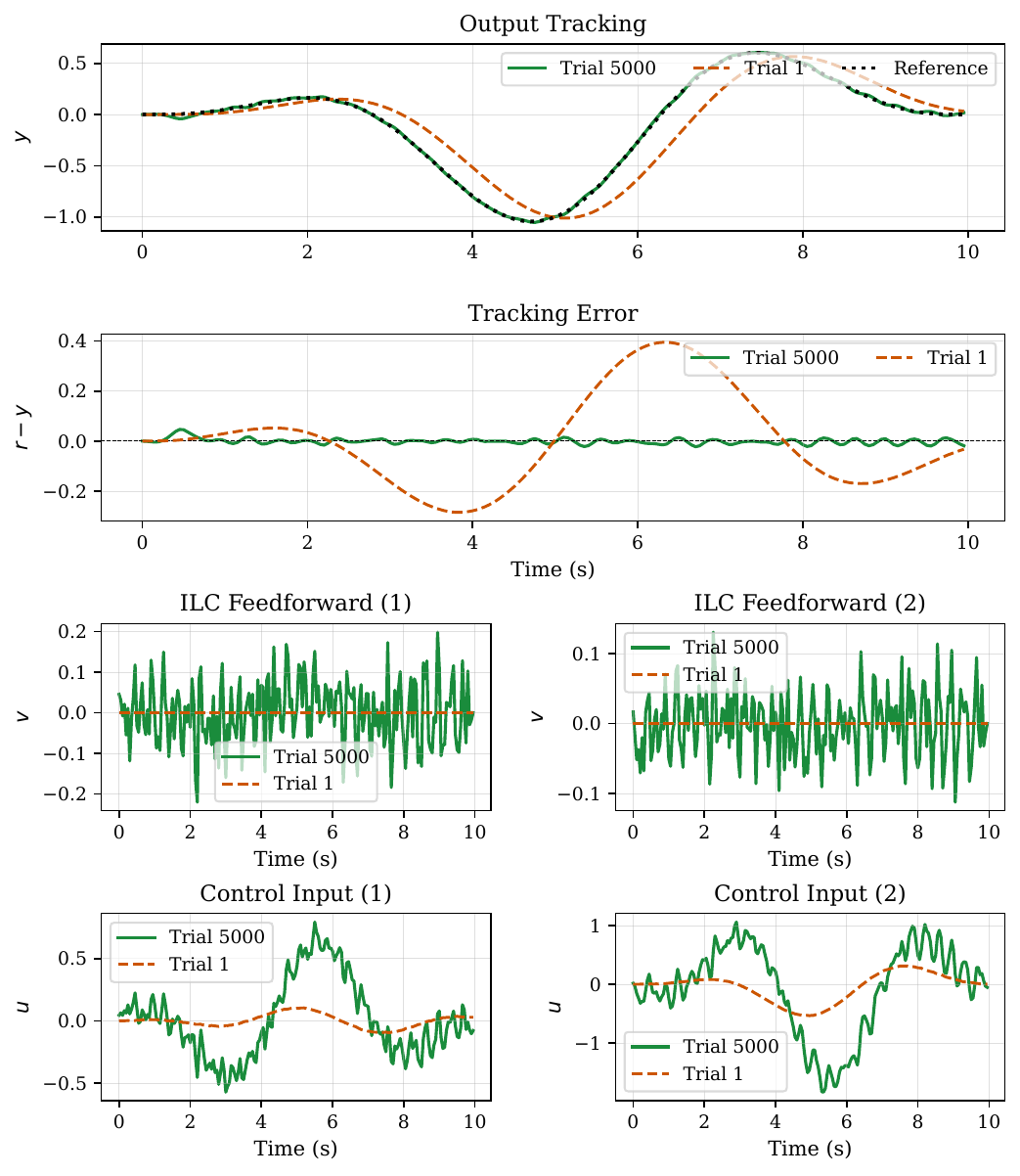}
\caption{\label{fig:with_reduction}
Encrypted ILC result for the reduced-coordinate update with range-space
reduction. The reference trajectory (dotted black line),
initial trial (dashed orange line), and final trial (solid green line) are shown.}
\vspace{-5mm}
\end{figure}

Fig.~\ref{fig:without_reduction} shows the full-coordinate update without
range-space reduction, included to illustrate null-space error accumulation.
Although the output trajectory at trial 5000 is closer to the reference than
at trial 1, the feedforward and plant inputs grow far beyond those of
Fig.~\ref{fig:with_reduction}, with the control input reaching up to ten
times larger in magnitude. This indicates that the $\ker(\Gcal)$-components
of the encryption-induced perturbation accumulate without correction.
By contrast, Fig.~\ref{fig:with_reduction}
shows that the reduced-coordinate update suppresses null-space error
accumulation while achieving tracking improvement.

\begin{table}[thb]
\centering
\caption{Encrypted ILC update time with non-packed and packed implementations.}
\label{tab:packing_time}
\begin{tabular}{c|cc}
\hline
 & Non-packed \eqref{eq:nonpack}
 & Packed \eqref{eq:packed}\\
\hline
Min [ms] & 15317 & 74 \\
Max [ms] & 17033 & 121 \\
Mean [ms] & 15609.3 & 105.3 \\
Std. dev. [ms] & 384.4 & 16.2 \\
\hline
\end{tabular}
\end{table}
For the computation time comparison, non-packed and packed reduced-coordinate
updates are executed over 200 trials, with results summarized in Table~\ref{tab:packing_time}.
The packed update requires \(105.3{\rm ms}\) on average, about \(148\)~times
faster than the \(15609.3{\rm ms}\) of the non-packed implementation,
consistent with the theoretical factor \(n_\Theta=199\) from the reduction
in external products.

%-----------------------------------------------------------------------
\section{CONCLUSION}

This paper proposed a Ring-LWE based encrypted ILC framework combining encrypted output feedback control 
based on the Ring-GSW external product, cloud-side trajectory storage, and ILC computation directly over encrypted data.
By formulating the ILC update in reduced coordinates, null-space error accumulation is
avoided, and distinct packing dimensions are used for the feedback and ILC phases. 
Numerical simulations confirmed that the range-space decomposition suppresses null-space error accumulation in the encrypted feedforward 
update, and that packing considerably reduces the update time. 
Future work includes deriving a formal bound on the encryption-induced error growth and experimental validation on a repetitive control platform.

%\section*{ACKNOWLEDGEMENT}
%
%This paper has been supported by NRF of Korea in 2021.

%%%%%%%%%%%%%%%%% BIBLIOGRAPHY IN THE LaTeX file !!!!! %%%%%%%%%%%%%%%%%%%%%%
{\renewcommand{\baselinestretch}{0.90}

}
\end{document}